\documentclass[prl,twocolumn,superscriptaddress,showpacs,amscd,amsmath,amssymb,verbatim]{revtex4}

\usepackage{graphicx,color}
\usepackage{textcomp}
\usepackage[OT1,T1]{fontenc}
\usepackage[squaren, Gray, cdot]{SIunits}
\usepackage{amsmath}
\usepackage{natbib}
\usepackage{revsymb}
\usepackage[colorlinks=true,linkcolor=blue,citecolor=blue,urlcolor=blue]{hyperref}
\usepackage{ulem}

\newcommand{\EIO}{Er$_2$Ir$_2$O$_7$}
\newcommand{\TIO}{Tb$_2$Ir$_2$O$_7$}
\newcommand{\co}{(Color online) }
\newcommand{\K}{~$\kelvin$}

\newcommand{\etal}{{\it et al.}}
\newcommand{\mK}{~$\milli\kelvin$}

\newcommand{\AIAO}{all-in/all-out}

\begin{document}

\title{Anisotropy tuned magnetic order in pyrochlore iridates}

\author{E. Lefran\c cois}
\email[Corresponding author: ]{lefrancois@ill.fr}
\affiliation{Institut Laue Langevin, CS 20156, 38042 Grenoble, France}
\affiliation{CNRS, Institut N\'eel, 38042 Grenoble, France}
\affiliation{Univ. Grenoble Alpes, Institut N\'eel, 38042 Grenoble, France}
\author{V. Simonet}
\affiliation{CNRS, Institut N\'eel, 38042 Grenoble, France}
\affiliation{Univ. Grenoble Alpes, Institut N\'eel, 38042 Grenoble, France}
\author{R. Ballou}
\affiliation{CNRS, Institut N\'eel, 38042 Grenoble, France}
\affiliation{Univ. Grenoble Alpes, Institut N\'eel, 38042 Grenoble, France}
\author{E. Lhotel}
\affiliation{CNRS, Institut N\'eel, 38042 Grenoble, France}
\affiliation{Univ. Grenoble Alpes, Institut N\'eel, 38042 Grenoble, France}
\author{A. Hadj-Azzem}
\affiliation{CNRS, Institut N\'eel, 38042 Grenoble, France}
\affiliation{Univ. Grenoble Alpes, Institut N\'eel, 38042 Grenoble, France}
\author{S. Kodjikian}
\affiliation{CNRS, Institut N\'eel, 38042 Grenoble, France}
\affiliation{Univ. Grenoble Alpes, Institut N\'eel, 38042 Grenoble, France}
\author{P. Lejay}
\affiliation{CNRS, Institut N\'eel, 38042 Grenoble, France}
\affiliation{Univ. Grenoble Alpes, Institut N\'eel, 38042 Grenoble, France}
\author{P. Manuel}
\affiliation{ISIS Facility, Rutherford Appleton Laboratory - STFC, OX11 0QX, Chilton, Didcot, United Kingdom}
\author{D. Khalyavin}
\affiliation{ISIS Facility, Rutherford Appleton Laboratory - STFC, OX11 0QX, Chilton, Didcot, United Kingdom}
\author{L. C. Chapon}
\affiliation{Institut Laue Langevin, CS 20156, 38042 Grenoble, France}
\date{\today}

\begin{abstract}

The magnetic behavior of polycrystalline samples of \EIO\ and \TIO\ pyrochlores is studied by magnetization measurements and neutron diffraction.
Both compounds undergo a magnetic transition at 140 and 130\K\ respectively, associated with an ordering of the Ir sublattice, signaled by thermomagnetic hysteresis. 
In \TIO, we show that the Ir molecular field leads the Tb magnetic moments to order below 40\K\ in the \AIAO\ magnetic arrangement. 
No sign of magnetic long range order on the Er sublattice is evidenced in \EIO\ down to 0.6 K  where a spin freezing is detected. 
These contrasting behaviors result from the competition between the Ir molecular field and the different single-ion anisotropy of the rare-earths on which it is acting. 
Additionally, this strongly supports the \AIAO\ iridium magnetic order. 

\end{abstract}


\maketitle


Attention of the condensed matter community was recently attracted by the iridates. 
Due to the interplay between a strong spin-orbit coupling, crystalline electric field (CEF) and moderate electronic interactions, the Ir$^{4+}$ ions could be close to a new spin-orbitronic state with Jeff = 1/2 \cite{BJKim2008,BJKim2009}, up to a local distorsion of the Ir$^{4+}$ environment \cite{Hozoi2014}. 
In the pyrochlores R$_2$Ir$_2$O$_7$ (R = rare-earth element), the Ir$^{4+}$ 5d electrons might then stabilize unprecedented electronic phases like Weyl semi-metal \cite{Balents2010,Wan2011,YBKim2012,Balents2014}.
Both the rare-earth and the Ir atoms lie on interpenetrated pyrochlore lattices where the corner-sharing tetrahedral arrangement may produce geometrical frustration. 
Moreover, these two sublattices might be magnetically coupled thus leading to novel magnetic behaviors. 
Almost all the members of the series exhibit a metal to insulator transition (MIT) when the temperature decreases, which would coincide with a magnetic transition \cite{Taira2001, Matsuhira2011}. 
The high-temperature electronic state and the transition temperature T$_{\textrm{MI}}$ both depend on the rare-earth. 
A macroscopic signature for the magnetic transition is a bifurcation in the zero-field-cooled (ZFC) and field-cooled (FC) magnetization at T$_{\textrm{MI}}$. 
It was argued from electronic structure calculations \cite{Wan2011} that the Ir sublattice orders at T$_{\textrm{MI}}$ in the antiferromagnetic "\AIAO" (AIAO) configuration (all magnetic moments pointing towards or away from the center of each tetrahedron). 
This non-collinear configuration can be stabilized by antisymmetric exchange interactions alone \cite{Elhajal2005,Arima2013}.
It is however difficult to probe due to the small value of the Ir$^{4+}$ magnetic moment and to the strong neutron absorption of Ir. 
In compounds with non magnetic R atoms (Eu, Y), so far only the results of resonant magnetic X-ray scattering \cite{Sagayama2013} and muon spin relaxation ($\mu$SR) assorted with strong hypotheses (e.g. absence of structural distortion) were interpreted as a direct probe of this magnetic configuration \cite{Sagayama2013,Zhao2011,Disseler2012,Disseler2014}.  

An alternative way to access the Ir$^{4+}$ magnetism is through the magnetic configuration of the rare-earth sublattice. 
The magnetic interactions between the rare-earth atoms being weak, this sublattice will be first sensitive to the Ir$^{4+}$ molecular field via the Ir-rare-earth coupling. 
Its magnetic response will depend on the Ir magnetic order, on the nature of the Ir-rare-earth coupling, and on the rare-earth magnetocrystalline anisotropy. 
An AIAO magnetic order of the Nd$^{3+}$ was for instance evidenced by neutron scattering in Nd$_2$Ir$_2$O$_7$ \cite{Tomiyasu2012}, compatible with the same magnetic configuration for the Ir sublattice. 
This result was however called into question by $\mu$SR studies \cite{Disseler2014}. 
To further investigate the nature of the Ir magnetic order and its coupling to the rare-earth sublattice, we focused our attention on Er$_2$Ir$_2$O$_7$ and Tb$_2$Ir$_2$O$_7$ \cite{Matsuhira2007}, displaying large magnetic moments on the rare-earth site and different types of anisotropy. 
In other pyrochlore families, like the titanates and stannates, the Er and Tb based compounds indeed show distinct magnetic ground states \cite{Gardner1999,Matsuhira2002,Sarte2011,Bondah-Jagalu2001,Mirebeau2005} and are thus expected to respond differently to the Ir molecular field. 

\par In this article, we present the magnetic properties of the pyrochlore iridates \EIO\ and \TIO\ probed by powder magnetization measurements and neutron diffraction. 
A radically different magnetic behavior of the rare-earth sublattice is evidenced: An AIAO order is observed on the Tb$^{3+}$ magnetic sublattice below 40 K whereas the Er$^{3+}$ sublattice does not order down to 600\mK. 
These results are discussed in connection with the rare-earth single-ion anisotropy (Tb$^{3+}$ easy-axis versus Er$^{3+}$ easy-plane) and with the magnetic coupling of the rare-earth with the Ir sublattice magnetism, about which we thus obtain relevant information. 


The pyrochlore iridates crystallize in the $Fd\bar3m$ cubic space group, with the O occupying the 48$f$ and 8$b$ Wyckoff positions, the rare-earth and the Ir occupying the 16$d$ and 16$c$ positions respectively \cite{Gardner2010}. 
Polycrystalline samples of R$_2$Ir$_2$O$_7$, with R = Tb$^{3+}$ (4f$^8$, $S$=3, $L$=3, $J$=6, $g_JJ$=9 $\mu_B$) and 
Er$^{3+}$ (4f$^{11}$, $S$=3/2, $L$=6, $J$=15/2, $g_JJ$=9 $\mu_B$), were synthesized, by solid-state reaction starting from the binary oxides and by a new flux method 
using CsCl as flux, for neutron diffraction and magnetometry measurements respectively. 
The structure and quality of the samples were checked by X-ray diffraction. 
The lattice parameter and the $x$ coordinate of the 48$f$ O were found at room temperature equal to 10.1606(2) \AA\ and 0.334(2) for the Er compound and equal to 10.2378(5) \AA\ and 0.35(2) for the Tb compound. 
A contamination by less than $\approx$ 2-3\% of Tb$_2$O$_3$ and Er$_2$O$_3$ parasitic phases was found in the Tb and Er samples used for the magnetometry measurements. 

\begin{figure}[h]
	\centering
	\includegraphics[scale=0.4,trim=7.2cm 0.9cm 0 0.3cm, clip]{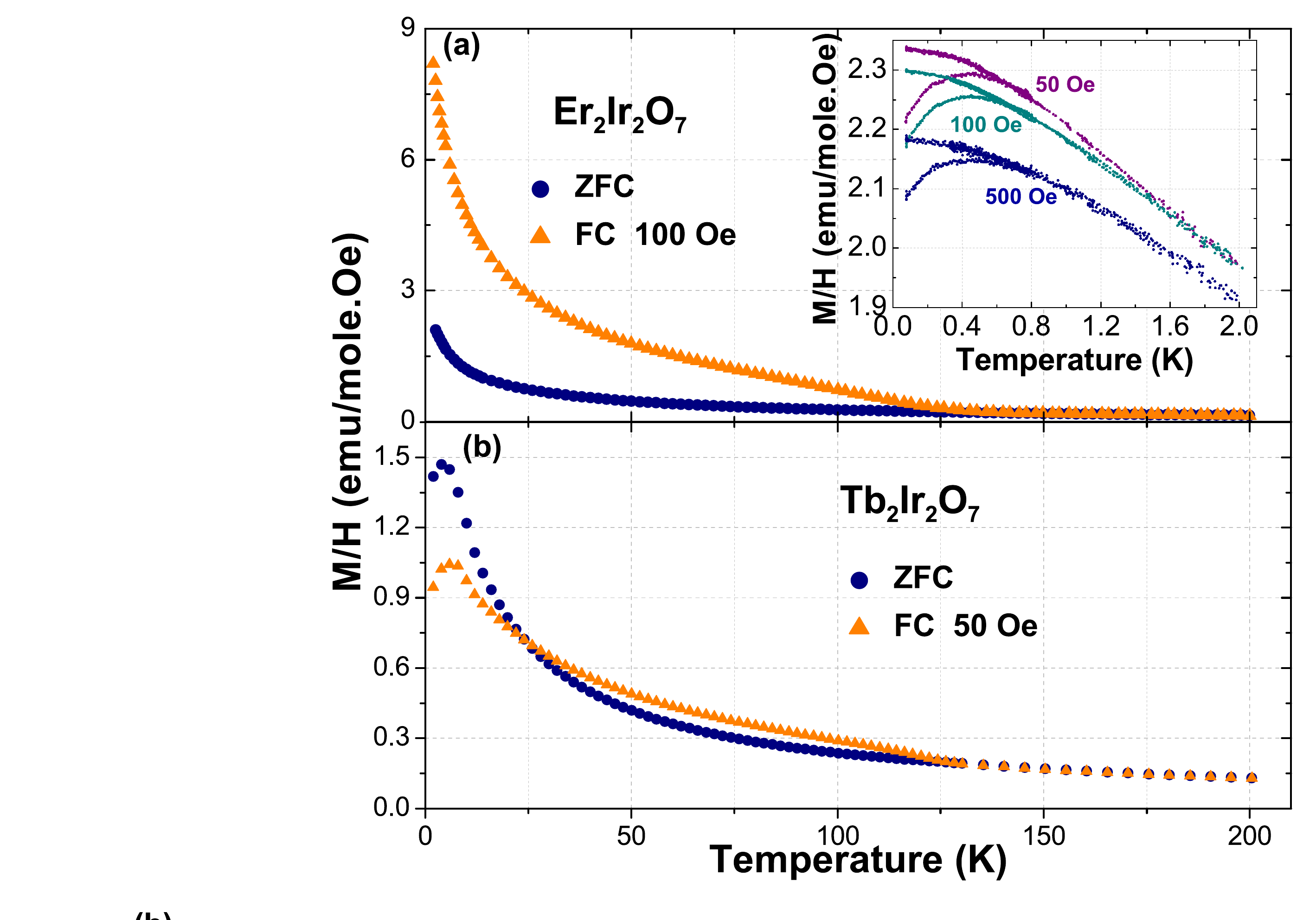}
	\caption{\co $M/H$ vs $T$ measured in 100 Oe in a ZFC-FC procedure for \EIO\ (a) and \TIO\ (b). The FC curves were measured while cooling. Inset: $M/H$ vs $T$ for \EIO\ measured between 0.08 and 2\K\ after ZFC and FC in different magnetic fields, the magnetic field in the FC procedure being applied below 4\K.} 
	\label{fig:Chi_T}
\end{figure}

The temperature and field dependence of magnetization ($M$) were measured, for both compounds using Quantum Design VSM and MPMS\textsuperscript{\textregistered} SQUID magnetometers down to 2 K, and down to 80\mK\ using a home-made SQUID magnetometer equipped with a dilution refrigerator \cite{Paulsen2001}. 
Neutron diffraction experiments were performed on powders i) at the Institut Laue-Langevin on the D7 diffractometer at 2\K\ and 50\K\ for \EIO\ and ii) at the ISIS facility for \TIO\, for which high-resolution data were collected on heating between 2 and 200\K, on the WISH diffractometer. 
Rietveld refinements for \TIO\ were carried out with the magnetic form factor of the Ir$^{4+}$ determined by Kobayashi \etal\  \cite{Kobayashi2011} using the FULLPROF program \cite{Rodriguez-Carvajal1993}. 

\begin{figure}[h]
	\centering
	\includegraphics[scale=0.4,trim=7.5cm 0 0 0, clip]{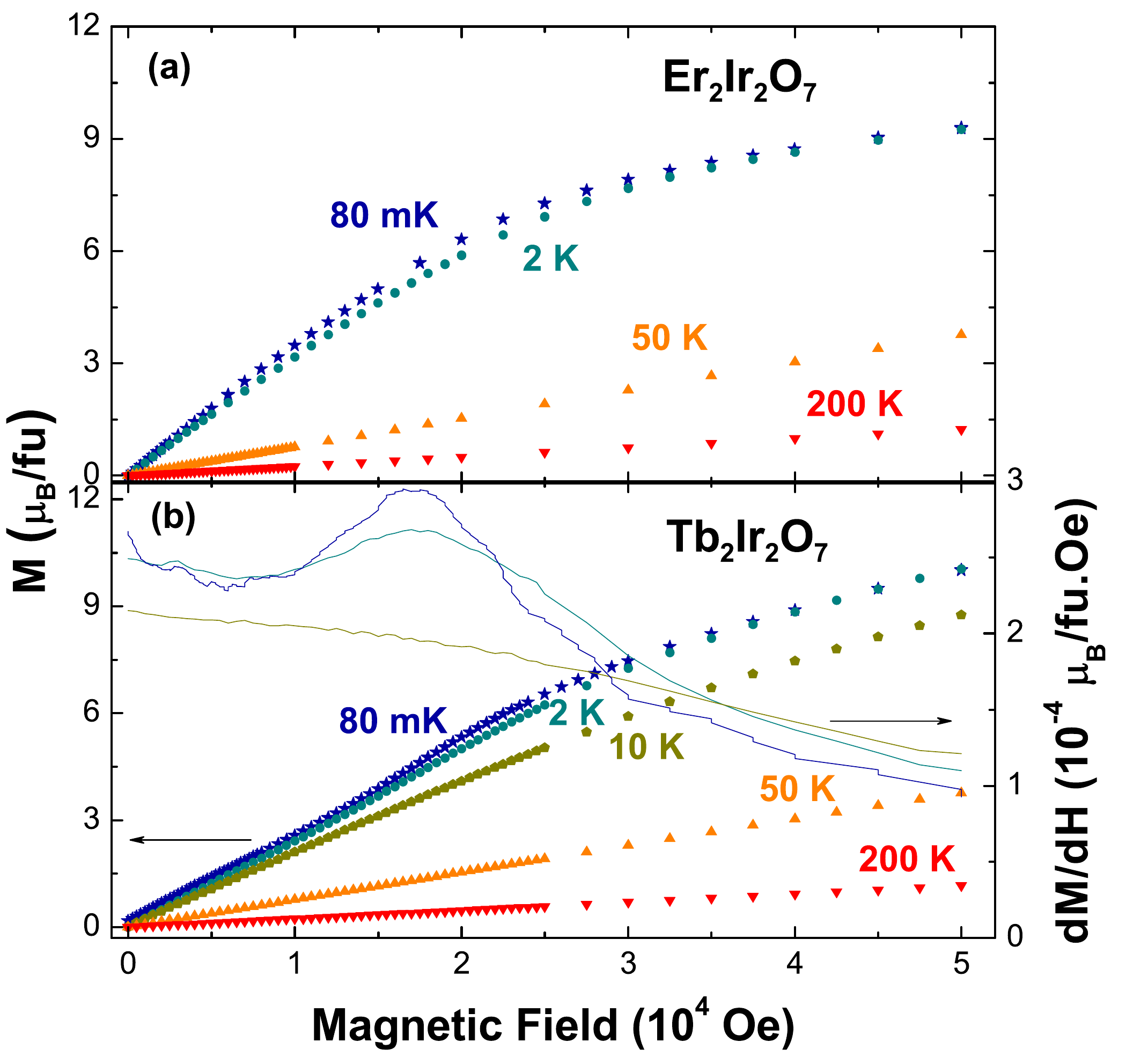}
	\caption{\co $M$ vs $H$ for \EIO\ (a) and \TIO\ (b) measured at different temperatures. Right side of (b): derivative of the magnetization curve for \TIO\ below 10\K\ showing a maximum indicative of a metamagnetic process.
	}
	\label{fig:M_H}
\end{figure}

For both compounds, the ZFC-FC magnetization was measured in an applied field of 100~Oe (see Fig.~\ref{fig:Chi_T}). 
A separation between the ZFC and FC curves is observed at 140 and 130\K\ for \EIO\ and \TIO\ respectively. 
This ZFC-FC behavior is consistent with previous results reported for this family 
showing that it coincides with the MIT \cite{Matsuhira2011,Shapiro2012}. 
Below this bifurcation, although the shape of the FC magnetization depends on the sample preparation, a general behavior is observed, summarized as follows: for \EIO, below T$_{\textrm{MI}}$, the FC magnetization remains above the ZFC one, both increasing down to 2\K\ without any sign of saturation. 
In \TIO, the two curves increase with decreasing temperature, the FC one lying above the ZFC one. 
Then at lower temperature, there is a crossing of the ZFC and FC curves, the former increasing faster than the FC one. 
Finally, both FC and ZFC curves in \TIO\ present a bump around 6\K. 
The isothermal magnetizations as a function of magnetic field are shown for \EIO\ and \TIO\ on Fig.~\ref{fig:M_H}. 
At the lowest temperature, a tendency towards saturation is observed in both compounds although not yet reached for the highest measured magnetic field of 80~kOe.
For \TIO, the magnetization additionally presents an inflection point, characteristic of a metamagnetic process at $\approx$ 18~kOe, which is absent above 10 K. 

\begin{figure}[h]
	\centering
	\includegraphics[scale=.3]{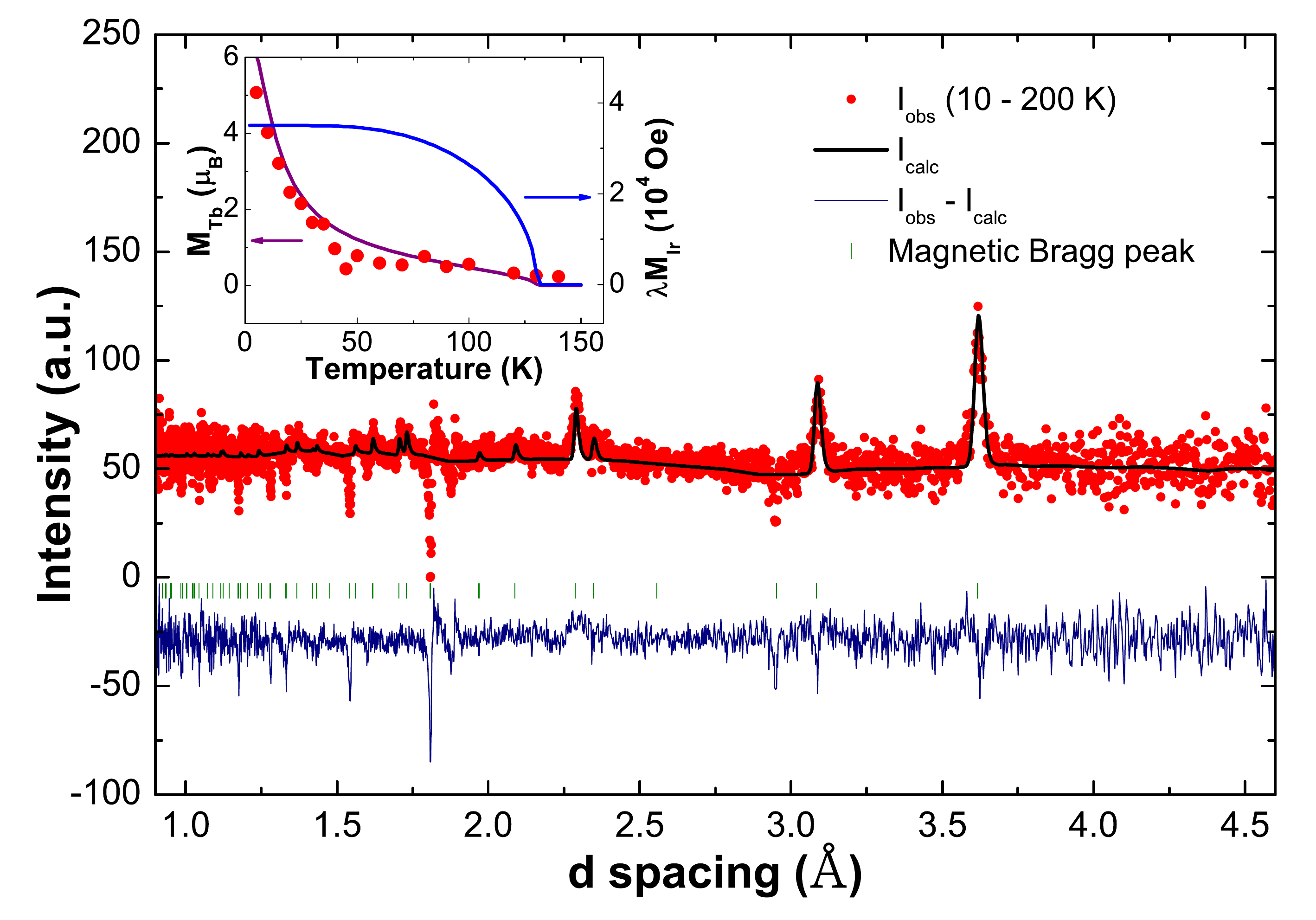}
	\caption{\co Difference between the 10 and 200\K\ neutron diffractograms recorded in \TIO\ (in red) and calculated intensity using the AIAO model for the Tb magnetic order (black line). Inset: Temperature dependence of the square root of the (2, 2, 0) magnetic reflection intensity (red dot). It is compared to the calculated Tb$^{3+}$ ordered moment (purple line) induced by the molecular field $\lambda M_{Ir}$ generated by the Ir$^{4+}$ magnetization whose temperature dependence is assumed to follow a Brillouin function (blue line).}
\label{fig:neutron_10K}
\end{figure}

\begin{figure}[h]
	\centering
	\includegraphics[scale=.35,trim=4.0cm 8.5cm 3.8cm 8cm,clip]{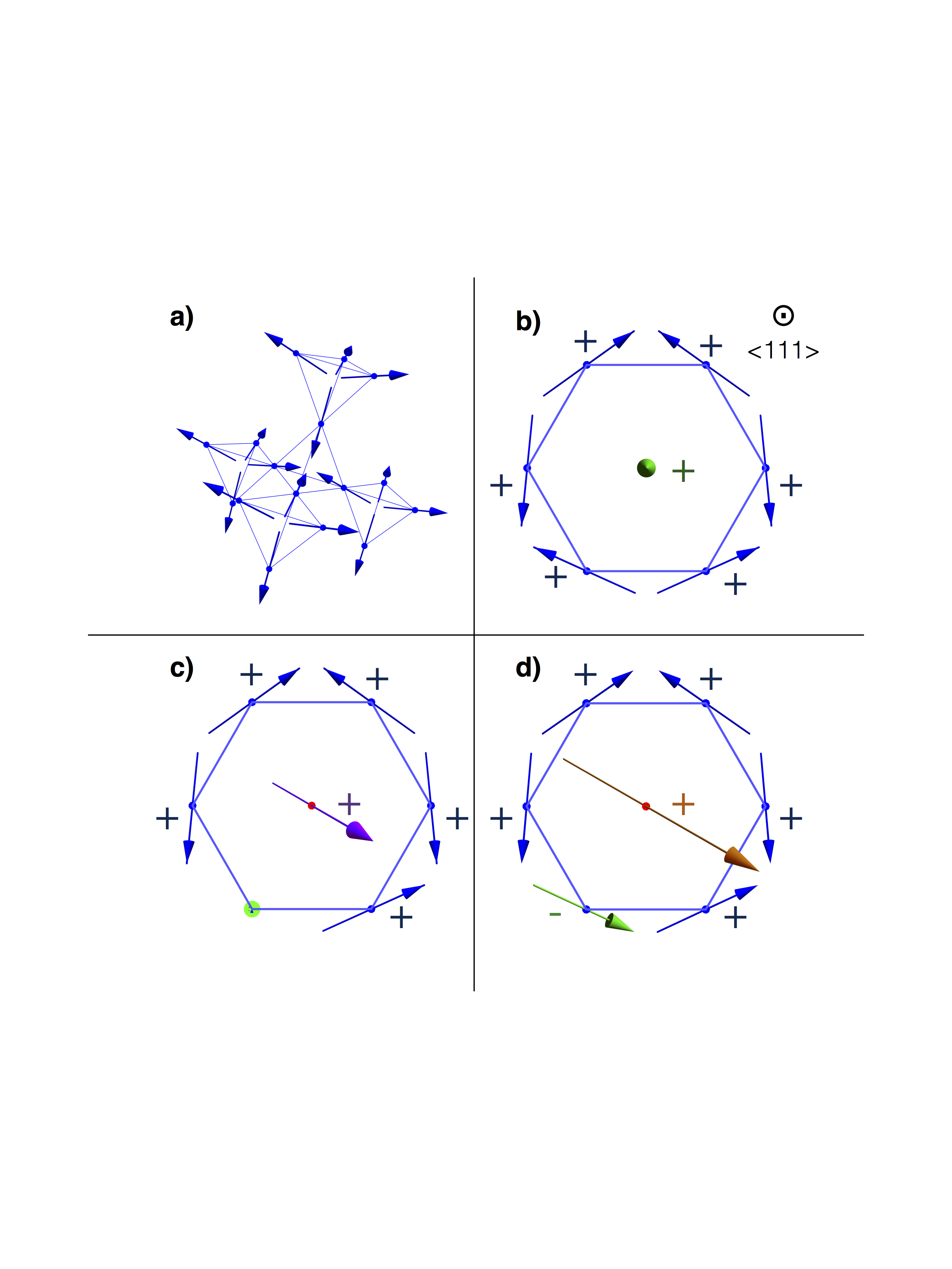}
	\caption{\co (a) AIAO magnetic configuration on the pyrochlore lattice. (b) The magnetic moments on a hexagon of 6 Ir$^{4+}$ ions in the AIAO configuration (blue arrows) yield a molecular field at the central Tb$^{3+}$ along the <111> cubic direction (green arrow). When adding a defect on one of the Ir site (in green), either a non magnetic ion (c) or a flipped magnetic moment (d), the Ir molecular field is tilted by an angle of respectively 29.5\degree\ (violet arrow) and 54.7\degree\ (orange arrow) with respect to the local <111> axis. The + (-) signs indicate the out-of-plane direction of the moments.}
\label{fig:Mag_struct_2}
\end{figure}

Neutron powder diffraction was performed on \TIO\ and \EIO. 
The difference between the diffractograms of \TIO\ at 200\K\ and at 10\K\ (above and below T$_{\textrm{MI}}$) reveals additional Bragg peaks indexable with the propagation vector {\bf k} = (0,~0,~0) (see Fig.~\ref{fig:neutron_10K}). 
From group theory and representation analysis \cite{Bertaut1968}, for the Ir 16$c$ site and the rare-earth 16$d$ site, the representation of the magnetic structure involves 4 irreducible representations (IR) among 10: $\Gamma~=~$GM$_2^+~+~$GM$_3^+~+~$GM$_5^+~+~2$GM$_4^+$ (notation of Miller-Love) \cite{Cracknell1979}, corresponding to the possible magnetic structures compatible with the $Fd\bar{3}m$ group symmetry. 
The Rietveld refinement of the neutron data shows that the AIAO magnetic configuration (GM$_2^+$ IR - shown in Fig. \ref{fig:Mag_struct_2}a) is the only one accounting correctly for the Tb magnetic ordering below 40 K and down to 2 K. 
The refined Tb$^{3+}$ magnetic moment at 10\K\ is $M(Tb) = 4.9 \pm 1~\mu_B $. 
The very weak magnetic moment at the Ir$^{4+}$ site could not be refined, being too small for the experimental sensitivity, i.e. lower than 0.2 $\mu_B$/Ir. 

The Tb$^{3+}$ ordered magnetic moment, proportional to the square root of the intensity of the \TIO\ magnetic Bragg peaks, starts to increase significantly below $\approx$ 40\K\ (see inset of Fig.~\ref{fig:neutron_10K}). 
Its temperature dependence down to 2\K\ does not follow a Brillouin function, as also reported for the Nd$^{3+}$ magnetic moment in Nd$_2$Ir$_2$O$_7$ \cite{Tomiyasu2012}. 
Its variation rather indicates that it is induced, through an effective Tb-Ir magnetic coupling, by the Ir molecular field, $\lambda \vec{M_{Ir}}$. 
To check this, we calculated the Tb$^{3+}$ induced magnetic moment by assuming a Brillouin function for the Ir$^{4+}$ magnetic moment temperature dependence and considering the following CEF model Hamiltonian:
\[
	\mathcal{H}_{\textrm{CEF}} = B^0_2 O^0_2 + B^0_4 O^0_4 + B^3_4 O^3_4 + B^0_6 O^0_6 + B^3_6 O^3_6 + B^6_6 O^6_6  
\]
where $O^m_n$ are Stevens' operators and $B^m_n$ are adjustable Stevens' parameters calculated from inelastic neutron scattering data for Nd$_2$Ir$_2$O$_7$ \cite{Watahiki2011}. 
Assuming that the environment is exactly the same as in Nd$_2$Ir$_2$O$_7$, we took for \TIO\ the Stevens' parameters 
\[ B^m_n(Tb)~=~\frac{B^m_n(Nd)}{\Theta_J(Nd)<r^n>_{Nd}}\Theta_J(Tb)<r^n>_{Tb} \]
where $\Theta_J = \alpha_J, \beta_J, \gamma_J$ stands respectively for the Stevens reduced matrix elements associated with $O^m_2$, $O^m_4$ and $O^m_6$ and where $<r^n>$ are radial integrals.
The Tb-Ir interaction is taken into account adding a term in the Hamiltonian 
$\mathcal{H}_{Tb-Ir}~=~\lambda \vec{M_{Ir}}(T)g_J\mu_B\vec{J}$.
The Tb$^{3+}$ magnetic moment is then computed as 
$\vec{M_{Tb}}~=~g_J\mu_B Tr( \vec{J} \exp(- \beta \mathcal{H}))$, where $\mathcal{H}~=~\mathcal{H}_{\textrm{CEF}}~+~\mathcal{H}_{Tb-Ir}$.
This model accounts well for the observed slow increase of $M_{Tb}$ below T$_{\textrm{MI}}$ which accelerates on lowering the temperature without any sign of saturation (see Inset Fig.~\ref{fig:neutron_10K}). 
It allows to extract a value for the Ir$^{4+}$ molecular field $\lambda M_{Ir}$, found $\approx$ 33 kOe at 10 K.

This is very different from the \EIO\ case, where no additional Bragg peak was observed down to 2\K\ in neutron powder diffraction (data not shown) indicating the absence of long-range magnetic ordering of the Er$^{3+}$ sublattice. 
Additional ZFC-FC magnetization measurements in various magnetic fields were performed down to 80 mK, in which the FC procedure started around 4\K\ (see inset of Fig. \ref{fig:Chi_T}). 
This allows to evidence thermomagnetic irreversibilities occurring below 4 K, hence not related to the Ir$^{4+}$ ordering. 
A ZFC-FC difference, most probably associated to some freezing of the Er$^{3+}$ magnetic moments, is observed around 600\mK.


The contrasting magnetic behavior of \EIO\ and \TIO\ can be understood by considering the difference of magnetocrystalline anisotropy between the two rare-earth ions. 
In the pyrochlore family, the O$^{2-}$ environment of the rare-earth is a distorted cube, constituted of puckered six-membered ring and two apical oxygens, which provides to this site a very pronounced axial symmetry along the <111> direction \cite{Gardner2010}. 
The sign of the $l$=2 main Stevens parameter, $B^{0}_2$, changes from negative for Tb$^{3+}$ to positive for Er$^{3+}$ ions, conferring to the former an axial anisotropy along the <111> direction and to the later a perpendicular easy-plane anisotropy. 
In the stannates and titanates, the magnetic behavior of the Tb and Er members is indeed in agreement with the easy-axis and easy-plane anisotropy respectively, hence with the sign of their $B^{0}_2$ term determined from neutron scattering \cite{Champion2003,Guitteny2013,Gardner1999,Mirebeau2005,Cao2009}.  

Assuming that the transition at T$_{\textrm{MI}}$ is second order as suggested by the absence of hysteresis in macroscopic measurements \cite{Ishikawa2012}, if the two Ir and rare-earth sublattices are coupled, they must order with the same IR. 
Our neutron diffraction results indicates that the Tb$^{3+}$ sublattice orders in the AIAO magnetic structure univocally given by the GM$_2^+$ IR. 
Moreover, as shown above, the temperature dependence of the integrated intensities supports an induced magnetic ordering, the Tb$^{3+}$ moments being polarized by the molecular field of the Ir$^{4+}$ sublattice in the AIAO arrangement. 
Figure \ref{fig:Mag_struct_2}b shows the net molecular field along the local <111> direction produced by the six first-neighbor Ir$^{4+}$ of each Tb site, assuming isotropic Tb-Ir exchange terms which are thus sufficient to induce the ordering of the Tb magnetic moments. 
On the contrary, the AIAO magnetic order of the Ir sublattice is incompatible with the easy-plane of anisotropy (magnetic configurations spanned by the GM$_3^+$ or GM$_5^+$ IR). 
Indeed, neither isotropic nor antisymmetric exchange terms between Ir and first neighbor Er can induce magnetic ordering, the latter producing terms in the Hamiltonian that cancel out when summed over all Ir neighbors on an hexagonal plaquette. 
For this reason, no induced magnetic moment is observed below $T_{\textrm{MI}}$ on the Er sublattice.

In addition to the Ir-rare-earth coupling, magnetic interactions between the rare-earth moments are expected to occur at lower temperature. 
The signature of these interactions could result in the features observed in the temperature and field dependence of the \EIO\ and \TIO\ magnetization below T$_{\textrm{MI}}$. 
In the Er$^{3+}$ case, this could be responsible for the onset of the freezing observed around 0.6\K, while in the Tb$^{3+}$ case it could explain the bump in the temperature dependence of the magnetization and the concomitant presence of a metamagnetic process below 6\K. 
The Tb$^{3+}$ sublattice could then first be polarized in the Ir$^{4+}$ molecular field yielding the AIAO arrangement, before it would feel, at low temperature, its own molecular field that has become dominant due to the large difference in the Ir$^{4+}$ and Tb$^{3+}$ ordered moments. 
This would lead to the same magnetic order (compatible with Tb~-~Tb antiferromagnetic interactions and easy-axis anisotropy), since no change is observed in the magnetic order by neutron scattering below 6\K.

Finally, we come back to the bifurcation observed between the ZFC and FC magnetizations starting at T$_{\textrm{MI}}$, which remains puzzling for a perfect AIAO antiferromagnetic arrangement. 
The same behavior is visible in another AIAO pyrochlore compound, Cd$_2$Os$_2$O$_7$ \cite{Sleight1974,Mandrus2001,Yamaura2012}.
The most probable explanation of this FC-ZFC characteristic behavior invokes intrinsic and/or extrinsic defects, present in these pyrochlore systems: an off-stoechimoetry can lead to the excess rare earth/Ir ions occupying the site of the counterpart ion, or to a Ir$^{5+}$/Ir$^{4+}$ substitution leading to a non magnetic site. 
The first type of defect has been shown to decrease the FC-ZFC difference \cite{Ishikawa2012}, whereas the second type increases this difference \cite{Zhu2014}. 
Another source of intrinsic defects comes from the presence of magnetic domains at 180$^{\circ}$ 
\cite{Arima2013,Tardif2014}. 
The magnetic moments located at the domain wall boundary have been shown to be free with respect to the nearest neighbor exchange coupling in another antiferromagnet with strong multi axial anisotropy \cite{Lhotel2011,Lhotel2013}. 
In  the pyrochlore iridates, the rare-earth ions feel the molecular field associated to the defective Ir$^{4+}$ neighborhood, and then be also polarized along the applied magnetic field. 
The molecular field of defective hexagons has a component along and perpendicular to the <111> directions, allowing the coupling of the Ir$^{4+}$ with the Er$^{3+}$ as well as the Tb$^{3+}$ ions (see Fig.~\ref{fig:Mag_struct_2}cd). 
In the Er case, the FC magnetization increases continuously and exhibits a shape compatible with an induced magnetization. 
In the Tb case, the onset of the Tb~-~Tb magnetic interactions at 6 K finally leads to a reversal of the FC polarized Tb$^{3+}$ moments that recover the antiferromagnetic AIAO structure and produces a global decrease of the FC magnetization with respect to the ZFC one.


In conclusion, our study of two members of the iridate pyrochlores gives a unified picture of the magnetic behavior of these materials, highlighting the strong coupling between the rare-earth and the Ir atoms. 
This coupling in \TIO\ allows us to establish the AIAO magnetic order of the Ir$^{4+}$ sublattice, which is thus expected to be a common feature of the family. 
The magnetic order of the rare-earth sublattice strongly depends on its magnetocrystalline anisotropy and on its compatibility with the Ir$^{4+}$ magnetic order. 
Beyond the interest in these systems for their potential topological non trivial states, these iridates thus provide an original playground to study novel magnetic properties in rare-earth pyrochlores submitted to a well-controlled $<111>$ local molecular field. 
This could allow to probe field-induced behaviors such as the quantum phase transition in easy-plane antiferromagnets \cite{Kraemer2012}.


We acknowledge D. Dufeu, Y. Deschanel and E. Eyraud for their technical assistance in the use of the magnetometers. 
We thank C. Paulsen for allowing us to use his SQUID dilution magnetometer, and A. Ralko and J. Robert for fruitful discussions.
We are thankful to G. Nilsen for his help for the experiment on D7.



\end{document}